\documentclass[12pt]{article}
\pdfoutput=1
\usepackage{graphics, color}
\usepackage{graphicx}
\usepackage{amsmath}
\usepackage{amssymb}
\usepackage{amsfonts}
\usepackage{tensor}
\usepackage[usenames,dvipsnames]{xcolor}
\usepackage[normalem]{ulem}
\usepackage[bottom]{footmisc}

\definecolor{orange}{rgb}{1,0.5,0}
\definecolor{grey}{rgb}{.5,.5,.5}
\definecolor{bluegreen}{rgb}{0,.5,.5}
\definecolor{darkgreen}{rgb}{0,.5,0}

\newcommand{\OL}[1]{ \hspace{.5pt}\overline{\hspace{-.5pt}#1
     \hspace{-.5pt}}\hspace{.5pt} }
\def\gsim{\, \rlap{$>$}{\lower 1.1ex\hbox{$\sim$}}\,}
\def\lsim{\, \rlap{$<$}{\lower 1.1ex\hbox{$\sim$}}\,}
\newcommand{\be}{\begin{equation}}
\newcommand{\ee}{\end{equation}}
\newcommand{\bea}{\begin{eqnarray}}
\newcommand{\eea}{\end{eqnarray}}

\begin{document}


\begin{titlepage}
\bigskip
\bigskip\bigskip\bigskip
\centerline{\Large \bf Brane/antibrane dynamics and KKLT stability}

\bigskip\bigskip\bigskip
\bigskip\bigskip\bigskip

 \centerline
 {\bf Joseph Polchinski\footnote{\tt joep@kitp.ucsb.edu }}
 \medskip

\centerline{\em Kavli Institute for Theoretical Physics}
\centerline{\em University of California}
\centerline{\em Santa Barbara, CA 93106-4030 USA}

\bigskip\bigskip\bigskip

\begin{abstract}
String theory has few or no stable nonsupersymmetric or de Sitter vacua, only metastable ones. Antibranes are a simple source of supersymmetry breaking, as in the KKLT model, but various arguments have been given that these  fail to produce the desired vacua. Proper analysis of the system requires identifying the correct effective field theories at various scales.  We find that it reproduces the KKLT conclusions.  This is an expanded version of a talk presented at SUSY 2015, Lake Tahoe.

\end{abstract}
\end{titlepage}

\baselineskip = 16pt
\tableofcontents

\baselineskip = 18pt

\setcounter{footnote}{0}


\section{Introduction}

In string theory, it seems that virtually all nonsupersymmetric vacua are ultimately unstable.  Essentially, everything not forbidden will eventually happen. Vacua have no conserved quantities, and nonsupersymmetric vacua are not protected by the energetic lower bounds arising from the supersymmetry algebra.\footnote{An early enunciation of this principle is~\cite{Fabinger:2000jd}.  Ref.~\cite{Giddings:2004vr} discusses one generic instability of de Sitter vacua, toward decompactification.}  This principle applies in particular to de Sitter vacua, as need to describe our current accelerating phase or an earlier inflating phase.

Of course, we live with metastability all the time.  If the nuclei in our bodies were allowed to minimize their energy, they would all reorganize into iron-56, or merge into a black hole, or decay via $\Delta B$ interactions.  Fortunately the rates for all of these are very small, and we do not spend much time worrying about them.  So perhaps it is not a fundamental problem that our own vacuum is metastable as well, but it makes things more challenging.  Metastability can be much harder to check than absolute stability (for example, it will hold only in limited regions of parameter space), and as in the example of the nuclei there may be many kinds of decay to consider.

There are now a variety of frameworks for constructing de Sitter vacua, including supercritical models~\cite{Silverstein:2001xn}, KKLT models~\cite{Kachru:2003aw}, large volume models~\cite{Balasubramanian:2005zx}, and negative curvature compactifications~\cite{Silverstein:2007ac}.  In this talk I will be discussing KKLT models.  This is not because of any prejudice about the likelihood of any particular scenario, but because the KKLT models involve a number of interesting dynamical issues.  These have come under scrutiny, and the construction has been challenged in a variety of ways.  Although most of the challenges are specific to the KKLT construction, it has sometimes been inferred that they represent a more general problem with the string landscape, or even that string theory cannot have realistic vacua.

I got caught up in this subject because of my interest in brane effective actions~\cite{Michel:2014lva}, but subsequently have had many discussions about other potential problems with KKLT models.  These models are complicated, but as with many problems in physics things become much clearer if one identifies the correct effective field theory. 

In \S2 I give a brief description of the KKLT models.  In \S3 I discussion issues with stability of antibranes.  In \S4 I discuss issues with the derivation of the effective four-dimensional description.  In \S5 I present some conclusions.

\section{The KKLT construction}

Let us start with a toy model of how all the de Sitter constructions work~\cite{Silverstein:2001xn,Kachru:2003aw,Balasubramanian:2005zx,Silverstein:2007ac,Silverstein:2004id}.  We will focus on just a single degree of freedom $r$, the radius of the compactified space; in a more complete treatment one would keep all of the K\"ahler moduli.  Consider first the compactification of M theory on a 7-sphere, with a nonzero flux on the sphere.
The effective four-dimensional potential in Planck units is
\be
V \sim \frac{N^2}{r^{21}} - \frac{1}{r^9} \,. 
\ee
The first term is from $N$ units of flux on the $S^7$, and the second (negative) term is from the positive curvature of the $S^7$.  This potential is shown in Fig.~1a. 
\begin{figure}[b]
\begin{center}
\vspace {-5pt}
\includegraphics[width=6.4 in]{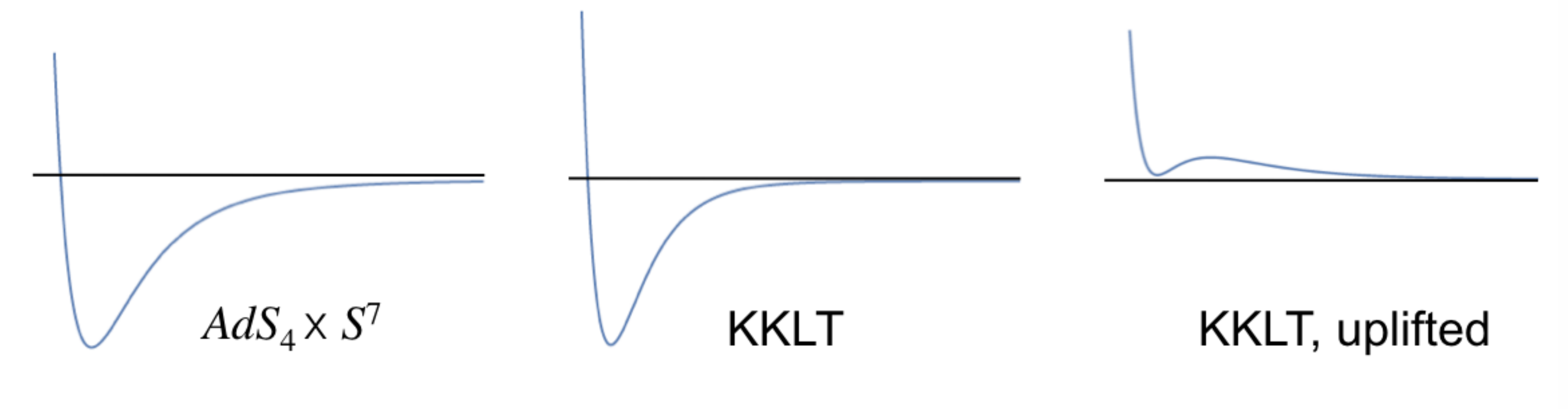}
\end{center}
\vspace {-10pt}
\caption{a) Potential for the $AdS_4 \times S^7$ model.  b) Potential for the KKLT model.  c) Potential for the KKLT model with antibrane.}
\label{fig:phase}
\end{figure}
 One sees that there is a stable supersymmetric $AdS_4 \times S^7$ minimum, and it is absolutely stable due to supersymmetry.  This gives one of the canonical examples of AdS/CFT duality~\cite{Maldacena:1997re}.

The KKLT construction~\cite{Kachru:2003aw} is similar in spirit.  There is a positive term from flux, partly canceled by the energy of O3-planes, but the curvature is zero and the negative term originates from a nonperturbative effect, either D-brane instantons or gaugino condensation on branes.  The K\"ahler potential and superpotential are
\be
K= - 3 \ln (\rho + \bar \rho)\,, \quad W = W_0 + A e^{-\rho} \,,
\ee
where $\rho = r^4 + i\chi$.  The potential is qualitatively similar (but goes more rapidly to zero at larger radius), and again produces a stable supersymmetric $AdS_4$ minimum.

The next step is to raise the energy by exciting the system.  In the KKLT vacua the supersymmetry is broken by adding an anti D3-brane;\footnote{Early work on antibrane SUSY breaking appears in~\cite{Antoniadis:1999xk}.} the metastability of this will be discussed in the next section.  The energy of the $\OL{\rm D3}$ can be adjusted by putting it in a warped throat, to give a final cosmological constant that is positive and/or close to zero.\footnote{As an aside, there is no known way to uplift $AdS_{D} \times S^k$ vacua, owing to the different behavior of the potential at large radius.  This 
 is why KKLT needed a more intricate construction.}

\section{Antibrane stability}

\subsection{Brane/flux annihilation, KPV, and singularities}

The `anti' in antibrane means that it has the opposite supersymmetry, and opposite charge, from other elements in the compactification.  In KKLT, the $\OL{\rm D3}$ has opposite D3 charge from that sourced by the background $H_3$ and $F_3$ fluxes through a Chern-Simons term.  Since everything not forbidden will eventually happen, these charges should find some way to annihilate.  This is the KPV process~\cite{Kachru:2002gs}, Fig.~2.
\begin{figure}[b]
\begin{center}
\vspace {-5pt}
\includegraphics[width=5.5 in]{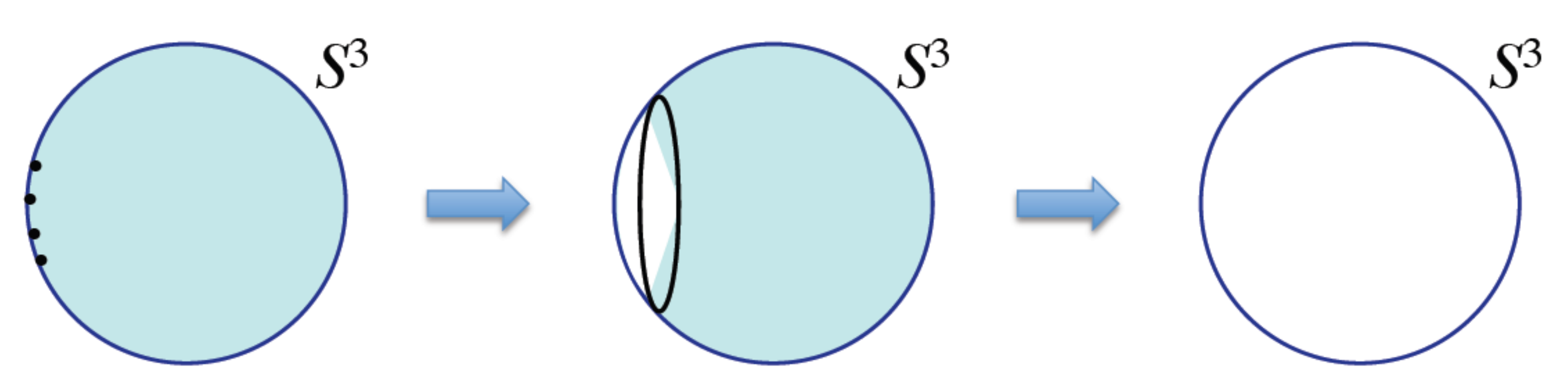}
\end{center}
\vspace {-10pt}
\caption{The KPV process a) Antibranes at the bottom of the KS throat, plus D3 charge carried by flux (shaded).  b) Antibranes polarized into an NS5, and flux partially neutralized.  c) Branes and flux annihilated.}
\label{fig:phase}
\end{figure}
The antibranes sit on an $S^3$ at the bottom of a Klebanov-Strasser throat~\cite{Klebanov:2000hb}, confined by a strong warp factor.  They polarize into an NS5-brane due to the presence of an $H_3$ flux on the $S^3$.  The NS5 can slip over the equator of the $S^3$ and then contract and disappear.  The net effect is that $M$ anti-D3's and $M$ units of D3 charge carried by flux disappear, where $M$ is the number of units of $H_3$ flux.  If there are fewer than $M$ anti-D3's to begin, some D3-branes will appear at the end.

This process is nonperturbative because there is a potential barrier for the NS5 to stretch over the equator, provided  that $p/M < 0.08$ where $p$ is the initial number of anti-D3's~\cite{Kachru:2002gs}.  One might imagine that a more local flux annihilation process is possible, but this is ruled out by the combination of the 5-form and 3-form Bianchi identities; this is discussed explicitly in~\cite{Michel:2014lva}.

\subsection{Brane effective field theory}

A challenge to this picture comes from the study of the backreaction of the antibranes on the background fields.  This leads to singularities whose physical admissibility has been questioned~\cite{McGuirk:2009xx,Bena:2009xk}, leading also to doubts about antibrane stability. 

In order to analyze this problem, we will start with the simplest case of a single $\OL{\rm D3}$, $p=1$.  D$k$-brane dynamics are described by the action
\be
S_{{\rm D}k} = -T_k \int d^{p+1}\xi\, e^{-\Phi} {\rm det}^{1/2}(G_{ab} + B_{ab} + 2\pi\alpha' F_{ab}) +  \mu_k \int {\rm Tr} \Bigl( B + e^{2\pi\alpha' F} \sum_q C_{q} \Bigr) \,. \label{braneact}
\ee
Here the metric $G$, the dilaton $\Phi$, and the form fields $B$ and $C$ live in the bulk, and the gauge field $F$ and embedding $X$ live on the brane.  This action has been used extensively, but its physical interpretation has always seemed a bit slippery.  The issue is that the brane sources the bulk fields, so these are singular at the brane and the action~(\ref{braneact}) diverges.

This is widely dealt with by working in the probe approximation, where the self-field is not inserted back into the action.  This is accurate for many purposes, but it is not a very physical approximation.  Formally one is taking the limit where the number of branes goes to zero, like a quenched approximation.  There are certainly many circumstances where the higher order terms are important.

A hint about the correct treatment comes from the work of Goldberger and Wise~\cite{Goldberger:2001tn}, who showed that codimension-two branes give rise to logarithmic divergences that can be summed via the renormalization group.  That is, there is a purely classical $\beta$-function.  This principal has also arisen in other contexts~\cite{Damour,Allen:1995rd}.  The point is that the RG is the right way to organize all effects by scale, not just quantum ones.

We should  extend this principal from the logarithmic divergences to all divergences. Regarding~(\ref{braneact}) as a low energy effective action~\cite{Weinberg:1978kz} is the right way to treat the classical divergences~\cite{Michel:2014lva}.  Expanding around flat spacetime, successively higher orders in the brane action are progressively more divergent, but these divergences are cut off at the UV string scale and are actually are smaller at higher order (provided $gp$ is small, where $g$ is the string coupling and $p$ the number of antibranes).  One then determines the finite value replacing these would-be divergences by matching onto the UV perturbative string theory.  Having said the right words, not much calculation is needed.\footnote{If there are flat directions, stability might depend on the signs of the higher terms, but for the single $\OL{\rm D3}$ the only flat direction is the motion along the bottom of the KS throat, and there must be minimum somewhere.}
  One can verify that the divergences from the effective action are the same as in the supergravity solution~\cite{Michel:2014lva}, but this is just dimensional analysis.  

For $p=1$, the only Lorentz-invariant degree of freedom in the low energy $\OL{\rm D3}$ action  is its position $X$.  All the $\OL{\rm D3}$ can do is move to the minimum of the potential, which would be some point near the bottom of the KS throat, and there is no possibility for any further rapid decay.  A single $\OL{\rm D3}$ is sufficient for the KKLT construction, at least in $O(1/10)$ of the examples, so we can say that as a point of principle that this is a closed issue.  The decay rate would be given by Eq.~(65) of~\cite{Kachru:2002gs}, which is highly suppressed for $p=1$. 
  For $2 < p < 1/g$ the story is similar, but one must do a calculation to see whether the antibranes separate or clump.  In the latter case they may lower their energy slightly by a small polarization, but in either case there is a stable final state.

The supergravity divergences would be seen if we tried instead to match on to supergravity in the UV, but this is the wrong UV theory for $gp < 1$.
In fairness to the antibrane instability literature, the supergravity analysis has been applied primarily for $gp > 1$.  The case $gp < 1$ was largely ignored in this literature prior to~\cite{Michel:2014lva}.

\subsection{Other issues}

The discussion above is sufficient as a proof of principle, but there are new issues that arise for $gp > 1$.  (KKLT~\cite{Kachru:2003aw} did not single out a particular range of $gp$).  There are a number of interesting technical results that might appear to point in the direction of KKLT instability.  There are many subtle issues, but in the end I do not see any results that indicate that these vacua are unstable.

\subsubsection{The tiny tachyon}

For $gp > 1$ the brane action becomes strongly coupled (if the antibranes are coincident), but the scale of the supergravity curvature involves the inverse of $gp$ and so supergravity may remain valid.  The supergravity singularity was argued to be resolved by polarization of the $\OL{\rm D3}$'s into an NS5~\cite{Kachru:2002gs}.  Ref.~\cite{Bena:2014jaa} finds that this polarized configuration is unstable toward separation of the $\OL{\rm D3}$'s.  This instability was termed a `giant tachyon', but the term `tiny tachyon' seems more appropriate because the lowest energy channel is precisely the one in which the antibranes do not blow up.

These calculations are intricate and I have not checked them.  If the result is correct, the KPV picture~\cite{Kachru:2002gs} is modified.  However, the SUSY breaking remains stable.  Once the $\OL{\rm D3}$'s separate, their only degrees of freedom are their positions, as in the discussion of $p=1$. They are moving in an effectively bounded space due to the warp factor, and so must find a minimum somewhere.  This final configuration is even more stable than the polarized one, because the barrier decreases with $p$.

\subsubsection{NS5 divergences}

For the brane action of the polarized system, the UV theory is not string theory but supergravity, because the NS5-brane is a field theory soliton.  This is an interesting matching problem~\cite{WIP}.  Ref.~\cite{Cohen-Maldonado:2015ssa} has argued for the existence of singularities based on the properties of a form $\cal B$, but we believe that this argument contains errors arising from the non-gauge invariance of $\cal B$.

\subsubsection{D6-branes}

Ref.~\cite{Bena:2012tx} shows that a $T$-dual configuration with $\OL{\rm D6}$'s does not polarize.  However, this does not reflect directly on the KKLT model, because the resolution of brane singularities is strongly dependent on dimensionality~\cite{Itzhaki:1998dd}.  In particular, for D$k$-branes with $k > 3$, the curvature always blows up near the brane, but the gauge theory on the branes becomes weakly coupled, and this is the correct description of the physics near the branes~\cite{Itzhaki:1998dd}.  The $\OL{\rm D6}$ result extends by naive $T$-duality to smeared $\OL{\rm D3}$'s~\cite{Bena:2012vz}, but this is not a physical configuration.

\subsubsection{Finite temperature}

It is shown in Refs.~\cite{Bena:2012ek,Bena:2013hr} that in certain cases black antibrane solutions do not exist (but see also~\cite{Hartnett:2015oda}). 
However, the definition of `antibrane'  is subtle: an anti-D$k$-brane was defined by a negative value of the flux
\be
\int_{S^{8-k}} \tilde F_{8-k} \,. \label{flux}
\ee
However, this `Maxwell' charge is carried by fluxes~\cite{Marolf:2000cb}, and the integral~(\ref{flux}) can change if flux flows through the $S^{8-k}$.
The actual conserved quantity (mod $M$) is the Page charge~\cite{Marolf:2000cb} 
\be
\int_{S^{8-k}} (\tilde F_{8-k}  - F_{6-k} \wedge B_2 )\,. \label{page}
\ee
It is this that keeps track of the antibrane charge when a horizon forms, and the no-go theorem no longer applies.

We should also note that a decay can be nonperturbatively slow at zero temperature and yet rapid at higher temperature, by passage over the potential barrier rather than through it.  The classic example is baryon number in the weak interaction~\cite{Kuzmin:1985mm}.  So the significance of a finite-temperature no-go theorem for zero temperature stability is not clear.

\subsubsection{A lower barrier?}

Various papers have suggested that the KPV process of Fig.~2 might proceed more rapidly by a different path with a lower barrier.  However, we believe that the suggested paths are unphysical.

To see one issue, consider the electrostatic energy
\be
\int d^dx \left(\frac12 (\nabla\phi)^2 + \rho\phi \right) \,.
\ee
Minimizing with respect to $\phi$ gives
$
\nabla^2 \phi = \rho ,
$
which is {\it not} Gauss's law.  Rather, Gauss's law $\nabla^2 \phi = -\rho$ is obtained by minimizing the electrostatic action
\be
\int d^dx \left(\frac12 (\nabla\phi)^2 - \rho\phi \right) \,.
\ee
It follows that the energy is not minimized by solving Gauss's law, and it is possible to reduce the energy below its physical value by taking an Ansatz for the electrostatic potential that does not satisfy Gauss's law.

Refs.~\cite{Blaback:2012nf,Danielsson:2014yga} consider an Ansatz for the electric part of the $B_6$ potential that does not satisfy Gauss's law, and find that the KPV potential energy can be reduced.  We see from the above discussion that this does not have physical significance.

Ref.~\cite{Danielsson:2015eqa} considers a different configuration.  As described in that reference this violates the Bianchi identity for the D3 charge.  This can be corrected by introducing additional fluxes, but these will carry additional energy that will almost certainly increase the potential above its original KPV value.

\section{From ten dimensions to four}

\subsection{From the antibrane to the nilpotent multiplet}

The antibrane is an object in ten-dimensional string theory, while the final analysis uses a four-dimensional effective description.  It has been argued that the latter is not correct~\cite{Brustein:2004xn,deAlwis,Sethi}.

To derive a low energy effective action, note first that the KKLT construction is based on two small numbers, $W_0$ and $e^{A_0}$.  We will retain the degrees of freedom that become massless when these parameters are taken to zero.  As $W_0 \to 0$ the sugra fields and K\"ahler moduli become massless.  As $e^{A_0} \to 0$ the fields at the bottom of the KS throat become massless.\footnote{Note that one is expanding around the limit of infinite warp factor in the KS throat.  Many studies of warped compactifications expand around the limit of constant warp factor, but one will not find the KKLT vacua in this regime.}  Thus the light degrees of freedom are
\be
{\rm EFT}_1: \quad\mbox{sugra + K\"ahler + throat bottom} \,.
\ee
The throat modes are still ten-dimensional, but conveniently we can use AdS/CFT duality to reexpress them in terms of a dual gauge theory~\cite{Klebanov:2000hb},
\be
{\rm EFT}_{1'}: \quad\mbox{sugra + K\"ahler + KS gauge theory} \,,
\ee
and all is four-dimensional.

Now, the KS gauge theory has metastable SUSY-breaking vacua, where there are antibranes at the bottom of the throat.  To see that this breaking is spontaneous, note that the KS gauge theory by itself is dual to a KS throat that is infinite in the UV direction.  Within this throat the KPV process~\cite{Kachru:2002gs} connects the state with antibranes to a supersymmetric state, so these are two states in the same field theory.  From another perspective, modes in the throat require two boundary conditions.  One comes in the IR, from the matching described in the previous section.   The second is imposed in the UV.  The two solutions there correspond to the Hamiltonian and the expectation value, so we can hold the Hamiltonian fixed and vary the state only.

In the gauge theory, the antibrane breaking is nonperturbative, and no simple description is known.  However, we only need the low energy effective theory.  The gauge theory is gapped aside from the goldstino~\cite{Volkov:1973ix} .  We need to describe the supersymmetric coupling of the goldstino to supergravity and the moduli.  Happily, the technology for this has been refined~\cite{Rocek:1978nb,Komargodski:2009rz}.  The goldstino can be described by a chiral multiplet $S$ which is nilpotent, $S^2 = 0$.  The nilpotence implies that the only degree of freedom is a fermion.  Then, as argued in Refs.~\cite{Ferrara:2014kva,Kallosh:2014wsa}, at very low energy we  have
\be
{\rm EFT}_2: \quad\mbox{sugra + K\"ahler + nilpotent multiplet} \,,
\ee
with 
\be
K= - 3 \ln (\rho + \bar \rho) + \OL SS\,, \quad W = W_0 + A e^{-\rho} + m^2S \,,
\ee
where $m$ is the KS breaking scale.
This gives the de Sitter minima of KKLT.\footnote{For inflating vacua such an effective field theory would apply during the period of slow roll, but higher energy degrees of freedom may be excited during reheating~\cite{deAlwis}.}   To good approximation, AdS energy from the sugra-$\rho$ sector adds to the $M^4$ from the KS sector.\footnote{A superpotential term $M^3$ might arise from the KS sector, but the effect of this is suppressed by the gravitational coupling.  Other couplings between sectors are similarly suppressed.}
Incidentally, for the case that the exponential superpotential term comes from gaugino condensation, there is a higher energy four-dimensional description EFT$_0$ including the gaugino condensation sector.  

The logic here seems simple and compelling, so where might it fail?


\subsection{Challenges}

\subsubsection{No-go theorems}

There is a long history of no-go theorems showing that supergravity and string theory do not have de Sitter solutions under certain conditions; Refs.~\cite{deWit:1986xg,Maldacena:2000mw,Giddings:2001yu,Kutasov:2015eba}  are some notable examples.  However, all known theorems are subject to certain assumptions, and none apply to the existing constructions~\cite{Silverstein:2001xn,Kachru:2003aw,Balasubramanian:2005zx,Silverstein:2007ac}.  For example,~\cite{deWit:1986xg,Maldacena:2000mw} omit string corrections, stringy sources (branes and O- planes), and quantum effects, and show that de Sitter solutions and also warped solutions are then forbidden, while~\cite{Giddings:2001yu} shows that these results survive the inclusion of some stringy sources but not others.   Ref.~\cite{Kutasov:2015eba} shows that in the heterotic string (where there are no stringy sources), this continues exactly in $\alpha'$, but not including quantum effects.

The fact that de Sitter vacua do not exist under these conditions is interesting, but not at all a concern.  The world is quantum mechanical; atoms would not be stable in a classical world either.  The classical limit is a boundary of moduli space, and the de Sitter vacua live in the interior.  From a rather different point of view, this was argued in~\cite{Dine:1985he}.

In summary, I see no direct or indirect limitation from the no-go theorems.

\subsubsection{Deriving the effective field theory}

It can be objected that the effective field theory of \S4.1 has been motivated but not derived.  Indeed, the wonder of effective field theory is that it makes hard problems much easier, and powerful results can be derived from simple reasoning.  Many important results that have been derived using effective field theory, notably nonrenormalization theorems and solutions of supersymmetric gauge theories, have never been shown directly from the UV theory. 

Refs.~\cite{Kallosh:2014wsa,Bergshoeff:2015jxa} have filled in some of the intermediate steps, but the KKLT model has many moving parts.  In fact, there is a fundamental limitation to deriving the effective field theory here: there is no well-defined starting point.  There is as yet no nonperturbative construction of string theory that would include these vacua (this has been emphasized in \cite{Banks:2003es}).  All studies of compactification are a patchwork of supergravity, string perturbation theory, and effective field theory.  
Arguably, these tools are being used within their regimes of validity, but the absence of a nonperturbative description of compactification is a notable lack of completeness in the current formulation of string theory. 

There are some objections to the use of effective field theory at all, but it is not clear why it should fail in this particular context.  In some cases it simply seems that the result is undesired.  A more principled objection~\cite{Banks:2003es} has been made based on intuitions as to the ultimate form of the theory of quantum gravity~\cite{Banks:2013fr}, which we do not share.  This last work also argues for the existence of stable de Sitter vacua.

\section{Conclusions}

Effective field theory is a powerful tool both for analyzing brane back-reaction, and for connecting the ten-dimensional picture with the four-dimensional one.  In the end, the original KKLT result has stood up well.  As far as we can see, none of the putative $10^{500}$ vacua has been eliminated.  Indeed, if the antibranes separate as suggested in \S3.3.1, they will remain stable at somewhat larger values of $p/M$, and perhaps there are $10^{501}$ vacua!

We would also like to focus attention on the need for a more complete nonperturbative construction of string theory, as mentioned at the end of the last section.   In a very different context, the inability of gauge/gravity duality to describe the black hole interior~\cite{Almheiri:2013hfa} is another sign that we need a nonperturbative description of gravitational bulk physics.  We are missing something, and it is bound to be interesting.

\section*{Acknowledgments}

I would like to thank Andrea Puhm for discussions of many aspects of this subject.  I  also  thank Iosif Bena, Ramy Brustein, Ulf Danielsson, Shanta de Alwis, Sergio Ferrara, Mariana Grana,  Shamit Kachru, Renata Kallosh, Sven Krippendorf, Juan Maldacena, Ben Michel, Eric Mintun, Fernando Quevedo, Philip Saad, Nati Seiberg, Sav Sethi, Eva Silverstein, Thomas van Riet, and Bert Vercnocke for discussions, and A. Puhm and R. Kallosh for comments on the draft.  This work was supported by NSF Grant PHY11-25915 (academic year) and PHY13-16748 (summer).

\end{document}